\documentclass[11pt]%,twoside]
{article} %%%%%%%%%%%%%%%%%%%%%%%%%%%%%%%%%%%%%%%%%%%%%%%%%%%%%%%%%%%%%%%%%%%%%%%%%%%%%%%%%%%%%%%%%%%%%
\usepackage{graphicx}
\usepackage{amsmath}
\usepackage{bbm}
\usepackage{cancel}
\usepackage[normalem]{ulem}
\usepackage{cite}

\usepackage{amsfonts}
\usepackage{amssymb}
\usepackage{amsmath}
\usepackage{latexsym}
\usepackage{xcolor}
\usepackage{url}
\usepackage{ntheorem}
\usepackage{braket}

\newcommand{\request}[1]{\mnote{change as suggested by referees}{\color{blue} #1}}
\renewcommand{\request}[1]{#1}
\usepackage{tikz}
\usepackage{pgfplots}
\usetikzlibrary{calc}

\usepackage{amssymb}
\usepackage[normalem]{ulem}
\usepackage[mathscr]{eucal}
\usepackage{url}
\usepackage{xcolor}
\usepackage{a4}
\usepackage{cite}
\usepackage{ntheorem}
\usepackage{graphicx}
\usepackage{psfrag}
\usepackage{subfigure}
\usepackage{latexsym}
\usepackage{amsmath}

\newcommand{\Ker}{\blue{\mathrm{Ker}}\,}
\newcommand{\Ima}{\blue{\mathrm{Im}}\,}

\newcommand{\Span}[1]{\blue{\mathrm{Span}}\,\{#1\}}

\newcommand{\kpotential}{\blue{\psi}}
\newcommand{\tpotential}{\blue{\chi}}
\newcommand{\hpotential}{\blue{\phi}}

\newcommand{\ptcheck}[1]{}%{\ptc{   checked up to here on #1} }

 \newcommand{\knotrace}{\red{\hat k}}
 \newcommand{\htrace}{\red{\gamma}}
 \newcommand{\htt}{\red{\hat h}}

\newcommand{\Koperator}{\red{Q}}

\DeclareFontFamily{OT1}{rsfs}{}
\DeclareFontShape{OT1}{rsfs}{m}{n}{ <-7> rsfs5 <7-10> rsfs7 <10->
rsfs10}{} \DeclareMathAlphabet{\mycal}{OT1}{rsfs}{m}{n}

\newcommand{\g}{$\bf \bar g^{\alpha\beta}$}

\newcommand{\blue}[1]{{\color{blue}#1}}
\newcommand{\red}[1]{{\color{red}#1}}

\newcounter{mnotecount}[section]

\renewcommand{\themnotecount}{\thesection.\arabic{mnotecount}}
\newcommand{\mnote}[1]%{}
{\protect{\stepcounter{mnotecount}}$^{\mbox{\footnotesize
$%\!\!\!\!\!\!\,
\bullet$\themnotecount}}$ \marginpar{%\color{red}%
\raggedright\tiny\em
$\!\!\!\!\!\!\,\bullet$\themnotecount: #1} }

\newtheorem{theorem}{\sc  Theorem\rm}[section]

\newtheorem{Theorem}[theorem]{\sc  Theorem\rm}

\newtheorem{corollary}[theorem]{\sc  Corollary\rm}

\newcommand{\eeal}[1]{\label{#1}\end{eqnarray}}
\newcommand{\bed}{\begin{deqarr}}
\newcommand{\eed}{\end{deqarr}}
\newcommand{\bedl}[1]{\begin{deqarr}\label{#1}}
\newcommand{\eedl}[2]{\arrlabel{#1}\label{#2}\end{deqarr}}

\newcommand{\mcU}{{\mycal U}}

\newcommand{\bel}[1]{\begin{equation}\label{#1}}
\newcommand{\bea}{\begin{eqnarray}}
\newcommand{\bean}{\begin{eqnarray}\nonumber}
\newcommand{\beal}[1]{\begin{eqnarray}\label{#1}}
\newcommand{\eea}{\end{eqnarray}}

\newcommand{\proof}{\noindent {\sc Proof:\ }}
\newcommand{\be}{\begin{equation}}
\newcommand{\eeq}{\end{equation}}
\newcommand{\ee}{\end{equation}}
\newcommand{\beqa}{\begin{eqnarray}}
\newcommand{\eeqa}{\end{eqnarray}}
\newcommand{\beqan}{\begin{eqnarray*}}
\newcommand{\eeqan}{\end{eqnarray*}}
\newcommand{\ba}{\begin{array}}
\newcommand{\ea}{\end{array}}

\newcommand{\R}{\mathbb R}

\newcommand{\eq}[1]{(\ref{#1})}

\newcommand{\beqar}{\begin{deqarr}}
\newcommand{\eeqar}{\end{deqarr}}

\newcommand{\beaa}{\begin{eqnarray*}}
\newcommand{\eeaa}{\end{eqnarray*}}

\newcommand{\tr}{\mbox{tr}}

{\catcode `\@=11 \global\let\AddToReset=\@addtoreset}
\AddToReset{equation}{section}

\renewcommand{\blue}[1]{#1}
\renewcommand{\red}[1]{#1}

\begin{document}
%\begin{titlepage}
%\begin{flushright}
%UWThPh-1996-37\\
%\today
%\end{flushright}
\title{On linearised vacuum constraint equations on Einstein manifolds. \protect\thanks{Preprint UWThPh-2020-02}}
\author{Robert Beig\thanks{
{\sc Email} \protect\url{robert.beig@univie.ac.at}},
Piotr T. Chru\'{s}ciel\thanks{
{\sc Email} \protect\url{piotr.chrusciel@univie.ac.at}, {\sc URL} \protect\url{homepage.univie.ac.at/piotr.chrusciel}}
\\
University of Vienna
}
\maketitle

\begin{abstract}
We show how to parameterise solutions of the general relativistic vector constraint equation on Einstein manifolds by unconstrained potentials. We provide a similar construction for the trace-free part of tensors satisfying the linearised scalar constraint.  \request{Previous work of ours has provided similar different constructions for solutions of the linearized constraints in the case where the cosmological constant $\Lambda$ is zero. We use our new potentials to show that one can shield linearised gravitational fields using linearised gravitational fields without imposing the TT gauge (as done in previous work), for any value of $\Lambda \in \R$.}
\end{abstract}

\tableofcontents

\section{Introduction}
 \label{s23XI16.1}

\request{The evidence, that Einstein's general relativity provides a very good model for our world, is overwhelming and keeps growing. Hence the obvious interest to achieve a detailed understanding of solutions of Einstein equations. Now, many such solutions can be constructed using the Cauchy problem. In general relativity the initial data for the physical fields are not arbitrary, but are required to satisfy a set of constraint equations. Proceeding in this manner, a key challenge is to obtain an exhaustive description of solutions
}
of the  constraint equations. As a step towards this, one might wish to obtain such a description in restricted settings, or  for simpler related equations.

For instance, consider the general relativistic vacuum vector constraint equation,
\begin{eqnarray}
 D_i(K^{ij} - \tr_g K g^{ij})
  =
   0
  \,.
  &
\eeal{18I20.1}
We will provide below an exhaustive description of solutions of this equation on three dimensional Einstein manifolds using \emph{unconstrained} potentials: We show (see Section~\ref{ss10I20.1} below)
 that for an arbitrary  symmetric tensor field $\kpotential_{ij}$ the tensor field $K=K_{ij}dx^idx^j$ defined by
\begin{equation}\label{18I20.2}
  K = \check  \Koperator \kpotential + \frac{1}{3 }\tr_g K g
  \,,
\end{equation}
satisfies \eqref{18I20.1}, where the operator $\check \Koperator$ is given by
\begin{eqnarray}
(\check \Koperator  \kpotential )_{ij}
 & =  &
  - 4 \Delta \kpotential _{ij} + 8 D_{(i} D^k \kpotential _{j)k} + \frac{4 R}{3} \kpotential _{ij} - \frac{8}{3} g_{ij} D^k D^l \kpotential _{kl}
   \nonumber
\\
 &&   +2 \Delta (tr_g \kpotential) g_{ij} - 2 D_ i D_j (tr_g \kpotential)
 - \frac{4R}{9} (tr_g \kpotential) g_{ij}
 \,,
 \label{added2}
\end{eqnarray}
and
\begin{equation}\label{24I20.1}
  \tr_g K = 2 D^\ell D^m \kpotential_{\ell m} + \frac{R}{3 }\tr_g
  \kpotential
   \,.
\end{equation}

Conversely  (see Corollary~\ref{C13XII19.3} below),  if a tensor field $K$ satisfies \eqref{18I20.1} on a three-dimensional simply connected Einstein manifold with vanishing second homology class, then there exists a tensor field $\kpotential $ such that  \request{$K$ is given by \eqref{18I20.2}}.

We present a similar result for the trace-free part of the linearised scalar constraint equation.
On an Einstein manifold, where
\begin{equation}\label{18I20.4}
  R_{ij}=\lambda g_{ij}
%  \,,
\end{equation}
(with the sign of $\lambda$ coinciding with that of the cosmological constant $\Lambda$),
the linearised scalar constraint becomes the following equation for a symmetric tensor field $h_{ij}$:
\begin{equation}\label{19I20.3}
-D^iD_i h^k{}_k + D^k D^\ell h_{k\ell} - \lambda  h^k{}_{k}= 0
 \,.
\end{equation}
For solutions with vanishing trace (e.g., because the trace has been gauged away by an appropriate choice of the initial data surface)
this becomes
\begin{equation}\label{19I20.4}
  D^k D^\ell h_{k\ell} = 0
  \,.
\end{equation}
It is not too difficult to check that for \emph{any} symmetric trace-free tensor field $\hpotential= \hpotential_{ij}dx^idx^j$ the tensor field $h=h_{ij}dx^i dx^j$ defined by the formula
\begin{equation}\label{19I20.5}
  h_{ij} = \epsilon_{i\ell k}D^\ell  \hpotential ^k{}_j
   +
    \epsilon_{j\ell k}D^\ell  \hpotential ^k{}_i
%    \,.
\end{equation}
is a symmetric trace-free tensor field solving \eqref{19I20.3}.
Under the same restrictions as for the vector constraint equation, we show (see Theorem~\ref{T13XII19.1} below)
that for  any trace-free tensor $h$ solving \eqref{19I20.4}  there exists a symmetric trace-free tensor field $\hpotential$ such that \eqref{19I20.5} holds.

In general gauges, where the trace of $h$ does not vanish, given one solution $\mathring h_{ij}$ of \eqref{19I20.3}, the remaining solutions are obtained by adding to $\mathring h$ a trace-free solution of \eqref{19I20.4}. So
our construction gives an exhaustive description of  solutions of \eqref{19I20.3} in this sense.

Recall that in~\cite{ChBeigTT}
(compare~\cite{JoudiouxShielding})
we provided a simple way of shielding gravity linearised at Minkowski space in transverse-traceless gauge (in a sense made precise by Theorem~\ref{T10I20.1} below), based on third-order unconstrained potentials for transverse-traceless tensors introduced in~\cite{BeigTT}. One of the motivations for the current work was to provide a shielding construction for linearised vacuum gravity which applies to initial data with a non-zero cosmological constant $\Lambda$, regardless of its sign and of the gauge. The analysis here applies to gravity linearised at Einstein metrics with $\Lambda \in \R$, while that in~\cite{ChBeigTT} works on any locally conformally flat manifolds but requires higher-order potentials.

An immediate corollary of our constructions below is the following shielding theorem:

\begin{Theorem}
\label{T10I20.1}
Let  $(h_{ij},k_{ij})$ be a smooth vacuum initial data set for the linearised gravitational field on a three-dimensional Einstein manifold $(M,g)$.
Consider open sets  $\mcU$, $\mcU'$ and $\Omega$ such that
$$
 \overline \mcU \subset \mcU' \subset \overline {\mcU'} \subset \Omega\subset M
 \,,
$$
and assume that \request{the cohomology groups $H_1$ and $H_2$ are trivial:}
$$
 H_1(\Omega)=H_2(\Omega)=\{0\}
  \,.
$$
%.
Then there exists a smooth vacuum initial data  set $(\tilde h_{ij},\tilde k_{ij})$, solution of the linearised vacuum constraint equations, which coincides  with $(h_{ij},k_{ij})$ on $\mcU$, and such that  $k$ and the trace-free part of $h$ vanish outside of $\mcU'$.
\end{Theorem}

In particular, for initial data for which %
%%`
%\begin{equation}\label{13I20.q}
%  g^{ij}h_{ij}=0
%  \,,
%\end{equation}
%%
%i.e., for which
the trace of $h$ can be gauged-away,
one obtains the screening of the full initial data set.

In the screening construction of~\cite{ChBeigTT} one needs first to transform the metric to a transverse-traceless gauge, which requires solving elliptic equations.
Neither  solving elliptic equations, nor applying preliminary gauge transformations,  is  needed in the approach taken here.

As discussed in~\cite{ChBeigTT}, a shielding construction provides immediately a gluing construction, the reader is referred to~\cite{ChBeigTT}
for details.

The proof of  Theorem~\ref{T10I20.1} is a repetition of the arguments given in \cite[Section~2.3]{ChBeigTT}, invoking instead the potentials of Theorems~\ref{T13XII19.1} and \ref{T13XII19.2}, and will be omitted.

The hypotheses above on $\Omega$ and the metric will clearly be satisfied if $\Omega$   is taken to be  a three-dimensional sphere with the standard round metric, or $\R^3$ with the flat or with the hyperbolic metric.

\section{Linearised constraint equations}
 \label{s16XII19.1}

Consider the three-dimensional general relativistic vacuum constraint equations,
\begin{eqnarray}\label{16XII19.21}
 &
 R(g) = 2 \Lambda + |K|^2_g - (\tr_g K)^2
 \,,
 &
\\
 &
 D_i(K^{ij} - \tr_g K g^{ij})
  =
   0
  \,.
  &
\eeal{16XII19.22}
Denoting by $h $ the linearisation of $g$  and by $k  $ the linearisation of $K$, the linearised version of \eqref{16XII19.21}-\eqref{16XII19.22} at a solution $(g,K\equiv 0)$ of the above reads
\begin{eqnarray}\label{16XII19.23}
 &
-D^iD_i h^k{}_k + D^k D^\ell h_{k\ell} - R^{k\ell}h_{k\ell}= 0
 \,,
 &
\\
 &
 D_i(k^{ij} - \tr_g k g^{ij})
  =
   0
  \,,
  &
\eeal{16XII19.44}
where $D$ denotes the covariant derivative of the metric $g$.

To fix notations, given a symmetric tensor $k_{ij}$, we denote by  $\knotrace$  its trace-free part:
\begin{equation}\label{17XII19.1}
  \knotrace_{ij}:= k_{ij} - \frac{1}{3} \tr_g k \, g_{ij}
  \,.
\end{equation}
In this notation, \eqref{16XII19.44} is equivalent to
\begin{eqnarray}\label{16XII19.31}
 &
 D_i(\knotrace^{ij} - \frac{2}{3}\tr_g k \, g^{ij})
  =
   0
  \,.
  &
\eea
This equation implies
\begin{equation}\label{16XII19.33}
  \epsilon^{\ell m j }D_m D_i \knotrace^{i}{}_{j} =0
  \,,
\end{equation}
and \eqref{16XII19.33} is equivalent to \eq{16XII19.44} on simply connected manifolds: In this case, if a trace-free tensor field $\knotrace$ satisfies \eqref{16XII19.33}, then there exists a function $\tau$ such that %
\begin{equation}\label{24I20.2}
 D_i \knotrace^{i}{}_{j} = D_j \tau
 \,.
\end{equation}
 The function $\tau$ is then uniquely defined by $\knotrace$ up to the addition of a constant.  It follows that $k_{ij}$ solves \eqref{16XII19.44} if and only if there exists a constant $c$ such that
\begin{equation}\label{17XII19.2}
  k_{ij}= \knotrace_{ij} + (\frac{\tau }{2} +c) g_{ij}
  \,,
\end{equation}
with the trace-free symmetric tensor $\hat k_{ij}$ solving \eqref{16XII19.33},
and with $\tau$ being a solution of \eqref{24I20.2}.

All solutions of \eqref{16XII19.33} are parameterised by   unconstrained symmetric tensor fields $\kpotential_{ij}$ in Corollary~\ref{C13XII19.3} below.

From now on we assume that $(M,g)$ is Einstein.

As already indicated, we will provide in Theorem~\ref{T13XII19.2} below an exhaustive description of the set of solutions of \eqref{16XII19.33} in terms of second-order potentials.

We pass now to a discussion of solutions of the scalar constraint equation \eqref{16XII19.23}.

While the following will not be assumed in our theorems below, one should keep in mind some
situations of particular interest, namely

\begin{enumerate}
  \item $M$ compact, or
  \item $(M,g)$ is the Euclidean space, or
  \item $(M,g)$ is the hyperbolic space.
\end{enumerate}

This allows any $\Lambda \in \R$.

If we denote by $\htrace$ the trace of $h$, we can write
\begin{equation}\label{16XII19.25}
  h_{ij} = %\hnotrace_{ij} +
  \htt_{ij} +\frac{1}{3} \htrace g_{ij}
  \,.
\end{equation}
The linearised scalar constraint equation \eqref{16XII19.23}  becomes
\begin{eqnarray}\label{16XII19.29}
 &
 (\frac 23  \Delta_g   +  \lambda) \htrace =  D_iD_j \htt^{ij}
 %D^k D^\ell \hnotrace_{k\ell}
  \,,
  &
\eea
\request{where $\lambda$ is as in \eqref{18I20.4}.}
This equation, viewed as a PDE for $\htrace$, has typically a finite dimensional set of solutions of interest. For instance, when the right-hand side vanishes the function $\tau$ will be zero on compact manifolds with $\Lambda<0$. Similarly
$\tau$  will be zero on a flat $\R^3$ and on hyperbolic space when restricting oneself to solutions which tend to zero at infinity, as easily follows afer applying the maximum principle for the Dirichlet problem on larger and larger balls. Compact manifolds with $\Lambda=0$ will lead to constants being the only solutions of the homogeneous equation. A finite dimensional set of non-trivial functions $\htrace$ might arise on some compact manifolds when $\Lambda >0$.
Note, finally, that under gauge transformations $h_{ij}\to h_{ij} + D_i \xi_j + D_j\xi_i$ we have $\htrace \mapsto \htrace + 2 D^i \xi_i$ so that a suitable choice of $\xi$ can bring $\gamma$ to a constant (possibly, but not necessarily, zero), but
\request{gauge choices is something that we wish to avoid,
as this typically requires solving further  equations which might be inconvenient, computationally expensive, physically irrelevant, and might require unwanted supplementary boundary conditions.
}

By linearity, it remains to describe the set of solutions of
\begin{equation}\label{19I20.1}
  D_iD_j \htt^{ij}=0
  \,,
\end{equation}
 this will be done in Theorem~\ref{T13XII19.1}.

\renewcommand{\g}{\red{$\bf \bar g g^{\alpha\beta}$}}
\section{Solutions of the linearised constraints on three-dimensional Einstein manifolds}

Let $(M,g_{ij})$ be a Riemannian 3-manifold which is locally conformally flat. Then (as pointed out in \cite{BeigTT}) the symmetric, trace-free tensor $H_{ij}$ given by the linearization of the Cotton tensor at the metric $g_{ij}$ in the direction of $h_{ij}$ is  3rd-order linear partial differential operator on $h_{ij}$, which is divergence-free and vanishes on $h_{ij}$'s which are conformal Killing forms of vectors $X_i$, i.e. %
$$
 h_{ij} = (L X)_{ij} = D_i X_j + D_j X_i - \frac{2}{3} g_{ij}\,\mathrm{div}_1 X
 \,.
$$
%.
We will restrict to $h_{ij}$ trace-free. Moreover we assume that $(M,g_{ij})$ is a space form, i.e. $R_{ijkl} = \frac 1 3 R g_{k[i}g_{j]l}$ with $R = \mathrm{const}$. It is then possible to write down a concise expression for  $t_{ij}$ as follows:
\begin{equation}
8 H[h] = \mathrm{rot}_2 \, (\mathrm{rot}_2^2   + L \,\mathrm{div}_2  - \frac{2 R}{3} ) h\,,
 \label{4XI19.0}
\end{equation}
where
$$(\mathrm{div}_2 \,h)_i = D_j h_i{}^j\,,
$$
 and where $\mathrm{rot}_2 \, h$ is the symmetric trace-free tensor given by
 $$(\mathrm{rot}_2 \, h)_{ij} = 2 \epsilon_{k\ell(i}D^k h^\ell{}_{j)}
 \,.
$$
%.

We note that the operator
$$
 \Koperator  := \mathrm{rot}_2^2  + L \,\mathrm{div}_2  - \frac{2 R}{3}
$$
appearing in \eqref{4XI19.0} is eight times the linearization of the trace-free Ricci tensor at $g_{ij}$ in the direction of trace-free tensors $h_{ij}$, namely
\begin{equation}\label{added}
(\Koperator  h)_{ij} = - 4 \Delta h_{ij} + 8 D_{(i} D^k h_{j)k} + \frac{4 R}{3} h_{ij} - \frac{8}{3} g_{ij} D^k D^\ell h_{k\ell} \,.
\end{equation}
Note also that $Q$ coincides with $\check  \Koperator$ defined in (\ref{added2}) when restricting to trace-free tensors. Last but not least, this is related to the linearization of the trace-free Ricci tensor at $g_{ij}$ as follows: If one now replaces the variation of the metric $h_{ij}$ by $h_{ij} + 1/2  \tr_g h   g_{ij}$ in that last tensor,   one obtains 1/8 of the hatted Q.

\subsection{Complexes}
 \label{ss10I20.1}

With the standard definition $(\mathrm{rot}_1 \,X)_i = \epsilon_i{}^{jk} D_j X_k$, and with $d$ denoting the differential \emph{of a scalar},  we have
\begin{eqnarray}
&
\mathrm{div}_2\, \mathrm{rot}_2 \, = \mathrm{rot}_1 \,\mathrm{div}_2\,,
\label{4XI19.1}
 &
 \\
 &
 (\mathrm{rot}_1)^2 + \mathrm{div}_2\,L =
\frac{4}{3}d \,\mathrm{div_1} + \frac{2 R}{3}
 \,,
&
\label{4XI19.2}
\\
&
 \mathrm{rot}_2 \, L = L \,\mathrm{rot}_1
\,.
\label{4XI19.3}
&
\end{eqnarray}
Indeed, \eqref{4XI19.1} and \eqref{4XI19.2} are direct computations, while \eqref{4XI19.3} follows from \eqref{4XI19.1} by noting that the formal adjoint $L^\dagger$ of $L$ is the negative of $\mathrm{div}_2$, and $ \mathrm{rot}_2^\dagger =  \mathrm{rot}_2$.

Using that $\mathrm{rot}_1d$ is zero one finds
\begin{equation}
\label{4XII19.6}
\mathrm{div}_2\,H[h] = 0
\,.
\end{equation}
Using formal adjoints we have
\begin{equation}\label{4XII19.4}
   (\mathrm{div}_2\,L)^\dagger   =  \mathrm{div}_2\,L
   \,,
\end{equation}
and a simple computation then gives
\begin{equation}\label{4XII19.5}
   H^\dagger   =  H
   \,.
\end{equation}
From this and \eqref{4XII19.6} one concludes that it also holds, for all vector fields $X$,
\begin{equation}
 \label{4XII19.7}
H[L X] = 0
 \,.
\end{equation}

Recall that a complex is called \emph{elliptic} if the sequence of symbols is exact.
We have  recovered the elliptic complex,
\begin{equation}
0 \rightarrow \Lambda_1 {\overset{L} \longrightarrow} S_0^2 {\overset{H} \longrightarrow} S_0^2 {\overset{\mathrm{div}_2} \longrightarrow} \Lambda_1 \rightarrow 0
 \,,
\end{equation}
which is a special case of the ``conformal complex'' derived in \cite{BeigTT} (compare~\cite{GasquiGoldschmidt,JoudiouxShielding}), which applies to any locally conformally flat metric.
Inspection of the identities above leads to the following  {elliptic} complex:
\begin{equation}\label{mom}
0 \rightarrow C^\infty {\overset{d} \rightarrow} \Lambda_1 {\overset{L \, \mathrm{rot}_1} \longrightarrow} S_0^2 {\overset{\Koperator } \longrightarrow} S_0^2 \,{\overset{\mathrm{rot}_1 \,\mathrm{div}_2} \longrightarrow}\, \Lambda_1 {\overset{\mathrm{div_1}} \longrightarrow} C^\infty \rightarrow 0
\,,
\end{equation}
where the composition of the fourth and fifth arrow is   $\mathrm{div}_2 \,H =0$, and the composition of the third and fourth arrow vanishes by taking formal adjoints:
\begin{equation}\label{6XII19.1--}
  (\Koperator \, L \, \mathrm{rot}_ 1)^\dagger =  - 2  \mathrm{rot}_1 \,\mathrm{div}_2 \,\Koperator =  - 2  \mathrm{div}_2\mathrm{rot}_2 \, \Koperator  = - 2  \mathrm{div}_2 \,H = 0
  \,.
\end{equation}
Ellipticity of \eqref{mom}, and of \eqref{ham} below, follows from the calculations in Appendix~\ref{A9I20.1}.

Similarly we have a third elliptic complex
\begin{equation}\label{ham}
0 \rightarrow C^\infty {\overset{L d} \longrightarrow} S_0^2 \,{\overset{\mathrm{rot}_2} \longrightarrow}\, S_0^2 \,{\overset{\mathrm{div}_1\mathrm{div}_2} \longrightarrow}\, C^\infty \rightarrow 0
\,.
\end{equation}
For this, from \eqref{4XI19.3} we find
\begin{equation}\label{4XIII19.8}
  \mathrm{rot}_2\, L \, d =    L \underbrace{\mathrm{rot}_1 \,d}_0 =0
  \,,
\end{equation}
and the composition of the last two arrows vanishes by taking formal adjoints.

The complex (\ref{mom}) will be called \emph{momentum complex}, since the condition on $h_{ij}$ to be the trace-free part of a field satisfying the linearized momentum constraints is exactly that $\mathrm{rot}_1 \, \mathrm{div}_2 \,[h]$ be zero. The complex in (\ref{ham}) is similarly related to the linearized Hamiltonian constraint, and will be referred to as the \emph{Hamiltonian complex}.

The complexes just discussed are collected, together with the de Rham complex , in the commutative  diagram of Figure~\ref{F19XII19.1}, where the composition of any two  horizontal or vertical  arrows vanishes.
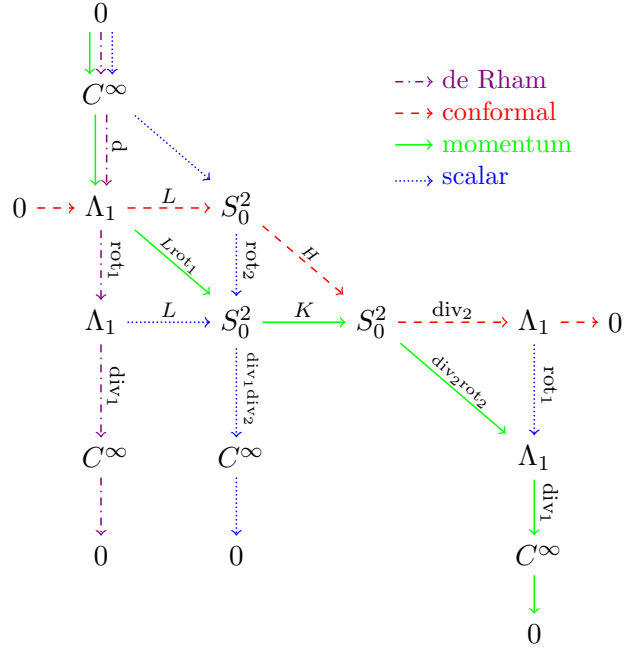
\begin{figure}
  \centering
\begin{tikzpicture}
\def\x{1.8} %% here, you set the x-distance between the boxes
\def\y{1.3} %% here, you set the y-distance between the boxes
\def\z{0.5} %% here, you set the shift between 4th and 5th row due to the div1div2 term
\def\l{0.24} %% specify the legend spacing
\def\d{0.22}

\node[align=center, name=04] at (0.4*\x,3*\y +\d) {$0$};

\node[align=center, name=11] at (\x,0*\y - \z) {$0$};
\node[align=center, name=12] at (\x,\y - \z) {$ \; C^\infty$};
\node[align=center, name=13] at (\x,2*\y) {$\Lambda_1$};
\node[align=center, name=14] at (\x,3*\y +\d) {$\Lambda_1$};
\node[align=center, name=15] at (\x,4*\y + 0.43) {$\;  C^\infty$};
\node[align=center, name=16] at (\x,5*\y +\d) {$0$};

\node[align=center, name=21] at (2*\x,0*\y - \z) {$0$};
\node[align=center, name=22] at (2*\x,\y - \z) {$\; C^\infty$};
\node[align=center, name=23] at (2*\x,2*\y) {$S_0^2$};
\node[align=center, name=24] at (2*\x,3*\y +\d) {$S_0^2$};

\node[align=center, name=33] at (3*\x,2*\y) {$S_0^2$};

\node[align=center, name=40] at (4.2*\x,-\y - \z + 0.2*\y) {$0$};
\node[align=center, name=41] at (4.2*\x,0*\y - \z) {$\; C^\infty$};
\node[align=center, name=42] at (4.2*\x,\y - \z) {$\Lambda_1$};
\node[align=center, name=43] at (4.2*\x,2*\y) {$\Lambda_1$};

\node[align=center, name=53] at (4.8*\x,2*\y) {$0$};

%redarrows
\draw[line width=0.2mm, dashed, ->,color=red] (04.east) -- (14.west);
\draw[line width=0.2mm, dashed, ->,color=red] (14.east) -- (24.west);
%\draw[line width=0.2mm, dashed, ->,color=red] (24.south east) -- (33.north west);  %% schraeger Pfeil wird haendisch ersetzt:
\draw[line width=0.2mm, dashed, ->,color=red] (3.95,3.88) -- (5.02,2.97);
\draw[line width=0.2mm, dashed, ->,color=red] (33.east) -- (43.west);
\draw[line width=0.2mm, dashed, ->,color=red] (43.east) -- (53.west);

%lilaarrows
\draw[line width=0.2mm, ->, dashdotted,color=violet] (16.south) -- (15.north);
\draw[line width=0.2mm, ->, dashdotted,color=violet] ($(15.south) + (0.075,0)$) -- ($(14.north) + (0.075,0)$);
\draw[line width=0.2mm, ->, dashdotted,color=violet] (14.south) -- (13.north);
\draw[line width=0.2mm, ->, dashdotted,color=violet] (13.south) -- (12.north);
\draw[line width=0.2mm, ->, dashdotted,color=violet] (12.south) -- (11.north);

%greenarrows
\draw[line width=0.2mm, ->,color=green] ($(16.south) - (0.15,0)$) -- ($(15.north) - (0.15,0)$);
\draw[line width=0.2mm, ->,color=green] ($(15.south) - (0.075,0)$) -- ($(14.north) - (0.075,0)$);
%\draw[line width=0.2mm, ->,color=green] (14.south east) -- (23.north west); %% schraeger Pfeil wird haendisch ersetzt:
\draw[line width=0.2mm, ->, color=green] (2.25,3.35 + 0.478) -- (3.25,2.492 + 0.478);
\draw[line width=0.2mm, ->,color=green] (23.east) -- (33.west);
%\draw[line width=0.2mm, ->,color=green] (33.south east) -- (42.north west);  %% schraeger Pfeil wird haendisch ersetzt:
\draw[line width=0.2mm, ->,color=green] (5.78,2.321) -- (7.18,1.12);
\draw[line width=0.2mm, ->,color=green] (42.south) -- (41.north);
\draw[line width=0.2mm, ->,color=green] (41.south) -- (40.north);

%bluearrows
\draw[line width=0.2mm, ->, densely dotted,color=blue] ($(16.south) + (0.15,0)$) -- ($(15.north) + (0.15,0)$);
%\draw[line width=0.2mm, ->, densely dotted,color=blue] (15.south east) -- (24.north west);  %% schraeger Pfeil wird haendisch ersetzt:
\draw[line width=0.2mm, ->, densely dotted,color=blue] (2.25,5.35) -- (3.25,4.492);
\draw[line width=0.2mm, ->, densely dotted,color=blue] (24.south) -- (23.north);
\draw[line width=0.2mm, ->, densely dotted,color=blue] (23.south) -- (22.north);
\draw[line width=0.2mm, ->, densely dotted,color=blue] (22.south) -- (21.north);
\draw[line width=0.2mm, ->, densely dotted,color=blue] (13.east) -- (23.west);
\draw[line width=0.2mm, ->, densely dotted,color=blue] (43.south) -- (42.north);

\scriptsize

\node[align=center, rotate=-90] at ($(14.north)!.5!(15.south) + (0.25,+0.05)$) {$\mathrm{d}$};
\node[align=center, rotate=-90] at ($(13.north)!.5!(14.south) + (0.2,+0.05)$) {rot$_1$};
\node[align=center, rotate=-90] at ($(12.north)!.5!(13.south) + (0.2,+0.05)$) {div$_1$};

\node[align=center, rotate=-90] at ($(23.north)!.5!(24.south) + (0.2,+0.05)$) {rot$_2$};

\node[align=center] at ($(14.east)!.5!(24.west) + (0,0.16)$) {$L$};
\node[align=center] at ($(13.east)!.5!(23.west) + (0,0.16)$) {$L$};
\node[align=center] at ($(23.east)!.5!(33.west) + (0,0.16)$) {$K$};
\node[align=center] at ($(33.east)!.5!(43.west) + (0,0.16)$) {div$_2$};

\node[align=center, rotate=-90] at ($(42.north)!.5!(43.south) + (0.2,+0.05)$) {rot$_1$};
\node[align=center, rotate=-90] at ($(41.north)!.5!(42.south) + (0.2,+0.05)$) {div$_1$};

\tiny
\node[align=center, rotate=-90] at ($(22.north)!.5!(23.south) + (0.2,+0.05)$) {div$_1$div$_2$};
\node[align=center, rotate=-40.63] at ($(6.48,1.7205)  + (0.1,0.0858)$) {div$_2$rot$_2$};
\node[align=center, rotate=-40.63] at ($(2.75,3.399)  + (0.1,0.0858)$) {$L$rot$_1$};
\node[align=center, rotate=-40.63] at ($(4.485,3.425)  + (0.1,0.0858)$) {$H$};

%%% the legend

\small

\node[align=left, color=violet, name=legend] at (4*\x,4*\y) {de Rham\\\color{red}conformal\\\color{green}momentum\\\color{blue}scalar};

\draw[line width=0.2mm, ->, dashdotted,color=violet] ($(legend.west) + (-0.5,2.5*\l)$) -- ($(legend.west) + (0,2.5*\l)$);
\draw[line width=0.2mm, dashed, ->,color=red] ($(legend.west) + (-0.5,.75*\l)$) -- ($(legend.west) + (0,.75*\l)$);
\draw[line width=0.2mm, ->,color=green] ($(legend.west) + (-0.5,-\l)$) -- ($(legend.west) + (0,-\l)$);
\draw[line width=0.2mm, ->, densely dotted,color=blue] ($(legend.west) + (-0.5,-3*\l)$) -- ($(legend.west) + (0,-3*\l)$);

\end{tikzpicture}
  \caption{Collected complexes.}\label{F19XII19.1}
\end{figure}

The question arises, whether the above complexes are exact at the second-to-last slots. This would provide an exhaustive description of solutions of the equations of interest. We establish this for the hamiltonian complex, cf. Theorem~\ref{T13XII19.1} below, but have not been able to for the momentum complex. In Theorem~\ref{T13XII19.2} below we will provide another complete classification of solutions of the momentum constraint. It would be of interest to determine exactness, or lack thereof, at this key slot,
of the momentum complex.

Let us mention that the momentum constraint equation on Minkowski spacetime is naturally related to the complex
$$
0 \rightarrow_0 \Lambda_1 \rightarrow_1 S^2 \rightarrow_2 S^2 \rightarrow_3 \Lambda_1 \rightarrow_0 0
 \,,
$$
%,
where
$\rightarrow_1$ is the Killing operator,
$\rightarrow_2$ is the linearized Schouten tensor,
$\rightarrow_3$ is the momentum operator.
Here $S^2$ denotes the space of all symmetric two-covariant tensors, without any trace condition.
This, however, is not a complex for  non-flat Einstein metrics. Note furthermore that when $g$ is the flat metric the above complex is closely related to the ``elasticity complex'' used in the area of finite elements methods for elasticity (see~\cite{Eastwood,ArnoldICM}, compare~\cite{PaulyZulehner}).

\subsection{The potentials}
 \label{ss11I20.1}

We are interested in the construction of solutions of the momentum
 and the scalar constraint,
 as well as of transverse-traceless tensors, on simply connected subsets $\Omega$ of three dimensional Einstein manifolds $(M,g)$. If $M$ itself is simply connected and complete, it follows from \cite[Corollary~2.4.10]{Wolf}
that $(M,g)$ is the three dimensional sphere, hyperbolic space, or Euclidean space. However, we neither assume completeness of $\Omega$, nor that of $M$, nor simple-connectedness of $M$.

We start with the
scalar constraint.
 We have:

\begin{Theorem}
\label{T13XII19.1}
Let $(M,g)$ be a three dimensional Einstein manifold with scalar curvature $R\in \R$.
On any open subset $\Omega$  of $M$ such that $H_1(\Omega)=\{0\}=H_2(\Omega)$,  the Hamiltonian complex \eqref{ham}
 is
 exact at the next-to-last entry. In other words,
a  trace-free and symmetric tensor field $t_{ij}$ on $\Omega$ satisfies
\begin{equation}\label{solve1}
  \mathrm{div}_1 \mathrm{div}_2 \,t = 0
%  \,.
\end{equation}
if and only if
there exists a symmetric trace-free two-covariant symmetric trace-free tensor field $\hpotential $
%and a covector field $\nu$
on $\Omega$  such that
%\ptcr{no control about boundary behaviour, right?} \bb{right}
%
\begin{equation}\label{13XII19.1a}
  \mathrm{rot}_2 \, \hpotential %+ L \, \mathrm{rot}_1 \,\nu
  = t
  \,.
\end{equation}
\end{Theorem}

{\sc\noindent Proof.}
The sufficiency has
been established when proving that \eqref{ham} is a complex. For necessity, note first that
Eq. (\ref{solve1}) and our assumptions on $\Omega$ imply  existence of a field $\tau_i$ such that
\begin{equation}
D^j t_{ij} = (\mathrm{rot}_1 \,\tau)_i
 \,.
\end{equation}
Thus the (non-symmetric) tensor given by $t_{ij} - \epsilon_{ijk} \tau^k$ is divergence-free with respect to the index $j$, i.e. $D^j(t_{ij} - \epsilon_{ijk} \tau^k) = 0$.

%Given a trace-free and symmetric tensor field $t_{ij}$  satisfying \eqref{solve1}
%we set
%
%\begin{equation}\label{9XII19.1}
%  \tau_{ijk} := \frac 12 \epsilon_{ij}{}^\ell t_{k\ell}
%  \,.
%\end{equation}
%
%Then $\tau_{ijk} = \tau_{[ij]k},\,\tau^i{}_{ji} = 0,\,\tau_{[ijk]} = 0$.
%
%For later use, we note the identity
%
%\begin{equation}\label{ident}
%g_{l[i}\tau_{jk]m}-g_{m[i}\tau_{jk]l} = 0
% \,.
%\end{equation}
%
%We also note that
%
%\begin{equation}\label{solve}
%  \mathrm{div}_1 \mathrm{div}_2 \,t = 0
%\end{equation}
%
%coincides with \ptc{multiplicative coefficient?}
%
%\begin{equation}\label{newab}
%  D_{[i}D^l\tau_{jk]l} = 0
%\,,
%\end{equation}
%
%and that  \eqref{solve1} implies \eqref{newab}.
%
It follows that the covector field
$$H_i = (t_{ij} - \epsilon_{ijk} \tau^k) \xi^j
\red{
 \,,
 }
$$
with $\xi^j$ given by
\begin{equation}
 \label{16IV20.1}
 \xi_i = D_i \xi
\,,
\end{equation}
where $\xi$ satisfies
\begin{equation}
D_i D_j \xi = - \frac{R}{6} g_{ij} \,\xi
 \,,
 \label{10I20.1}
\end{equation}
%
%\eqref{16IV20.1}-\eqref{10I20.1},
is divergence-free, and subsequently there exists a tensor field $H_{ij} = H_{[ij]}$ such that
\begin{equation}\label{Hij}
D^j H_{ij} = H_i
 \,.
\end{equation}
%

%We continue by considering the set of vectors of the form
%%
%\begin{equation}
% \label{16IV20.1}
% \xi_i = D_i \xi
%\,,
%\end{equation}
%%
%}
%with $\xi$ satisfying
%%
%\begin{equation}
%D_i D_j \xi = - \frac{R}{6} g_{ij} \,\xi
% \,.
% \label{10I20.1}
%\end{equation}
%%
Some comments on the set of such $\xi$'s are in order.
When $\Omega$ is connected and simply connected,   the set of solutions $\xi$ of this equation forms a four dimensional vector space. The solutions of \eqref{10I20.1} are then determined uniquely throughout $\Omega$ by the values of $\xi$ and $D \xi$ at one point. On a flat $\R^3$ the $\xi_i$'s are the parallel vectors, forming a three-dimensional vector space, and are uniquely defined everywhere by the value of $\xi_i$ at one point.
Both these claims can be established
%remains true for any simply connected $\Omega$ in a three-dimensional Einstein manifolds
by usual globalisation arguments. (Compare~\cite{Nomizu} for a proof of a similar statement for the Killing vector equation; identical arguments apply to \eqref{10I20.1}, see also \cite{HallExtension}.
Here it is useful to keep in mind that a Riemannian Einstein metric is real-analytic in harmonic coordinates so that, without loss of generality, we can assume that $(M,g)$ is analytic.)

 If $\Omega$ is a connected subset of the sphere or hyperbolic space,  and when the sphere or the hyperbolic space are embedded in a standard way in $\R^4$, the $\xi$'s are obtained by restricting to $\Omega$ the linear functions on $\R^4$;  on a sphere, these are the  $\ell= 1$ -- spherical harmonics.

Returning to our main line of thought, let
us show that any solution
  $H_{ij}$  of (\ref{Hij}) can be written in the form
\begin{equation}\label{V}
H_{ij} = V_{ijk} \xi^k + M_{ij} \xi\,,
\end{equation}
with smooth tensor fields $V_{ijk} = V_{[ij]k}$ and $M_{ij} = M_{[ij]}$,
with $M_{ij}\equiv 0$ when $(M,g)$ is flat.
 To see this, we note that the field $H_{ij}$ defines a  linear map on the space of $\xi$'s and thus can be written in the form
\begin{equation}
H_{ij} (x) = V_{ijk}(x) \xi^k(N)  + M_{ij}(x)  \xi (N)
\,,
 \label{10I20.11}
\end{equation}
where $N$ is some arbitrarily chosen point on $M$. Likewise there exist  smooth fields such that
\begin{equation}
\xi^j(x) =B^{j}{}_{k}(x) \xi^k(N)  + B^j(x)  \xi (N)
\,,
 \quad
\xi (x) =E _{k}(x) \xi^k(N)  + E (x)  \xi (N)
\,,
\label{9XII19.3}
\end{equation}
with $B^i\equiv 0$ and $E\equiv 0$ when $(M,g)$ is flat.
It is not difficult to check that \eqref{9XII19.3} can be smoothly inverted to give
\begin{equation}
\xi^j(N) =C^{j}{}_{k}(x) \xi^k(x)  +  C^j(x)  \xi (x)
\,,
 \quad
\xi (N) =F _{k}(x) \xi^k(N)  + F (x)  \xi (N)
\,,
\label{9XII19.4}
\end{equation}
with $C^i\equiv 0$ and $F\equiv 0$ when $(M,g)$ is flat.
Inserting (\ref{V}) into (\ref{Hij}) gives
\begin{equation}
D^j(V_{ijk} \xi^k + M_{ij} \xi) = (t_{ij} - \epsilon_{ijk} \tau^k) \xi^j
\end{equation}
%Equation~\eqref{9XII19.6} inserted in \eqref{xi} gives
%
%\begin{equation}\label{omega}
%  \big(
%   - \frac{R}{6} U_{[ji]} \xi + (D_{[i} U_{j]k}) \xi^k + (D_{[i} V_{j]}) \xi - V_{[i} \xi_{j]}
%     \big)
%       = \tau_{ijk} \xi^k +  \xi_{[i} \tau_{j]}\,.
%\end{equation}
%
%\ptcheck{ 9XII19}
If $R\ne 0$, at every point $p$ we can find a function $\xi$ which equals zero at $p$ and has arbitrary gradient there. It follows that
\begin{equation}\label{anti}
D^j V_{ijk} + M_{ik} = t_{ik} - \epsilon_{ikl} \tau^l
\,.
\end{equation}
Similarly one sees that
\begin{equation}
- \frac{R}{6} V_{ij}{}^j + D^j M_{ij} = 0\,,
\end{equation}
but we will not use this equation.
An analogous argument applies when $R=0$.

One easily checks that
\begin{equation}
V_{(i|j|k)} = V_{i[jk]} + V_{k[ji]}
\end{equation}
%A similar argument gives
%
%\begin{equation}\label{main}
%2( D_{[i} U_{j]k} + g_{k[i} V_{j]}) = \tau_{ijk} + g_{k[i} \tau_{j]}\,.
%\end{equation}
%
%Now define
%
%$$ h_{ij}:= U_{ij}- \omega_{ij} -  \frac{1}{3}U_l{}^l g_{ij}
% \,,
%$$
%,
%where $\omega_{ij} = U_{[ij]}$. Thus $h_{ij}$ is symmetric and trace-free. Inserting this into (\ref{main}) we find
%
Thus, from (\ref{anti}), we obtain
\begin{equation}\label{symm}
D^j(V_{i[jk]} + V_{k[ji]}) = t_{ik}
\end{equation}
We can write $V_{i[jk]}$ as $V_{i[jk]} = - \epsilon_{jk}{}^\ell A_{i\ell}$. Next we can decompose $A_{ij}$ as $A_{ij} = a_{ij} + \frac{1}{3} g_{ij} A_\ell{}^\ell + \omega_{ij}$, where
$a_{ij} = a_{(ij)}$ and $\omega_{ij}$ is antisymmetric. Inserting into (\ref{symm}) shows that $A_\ell{}^\ell$ does not contribute and, using that $t_{ij}$ is trace-free, that
$D_{[i} \omega_{jk]} = 0$.  Consequently
\begin{equation}
t_{ik} = (\mathrm{rot}_2 \, a)_{ik} + (L \, \mathrm{rot}_1 \,\nu)_{ik}\,,
\end{equation}
where $\omega_{ij} = 2 D_{[i} \nu_{j]}$.
Since $L \, \mathrm{rot}_1= \mathrm{rot}_2 \, L$, setting
$$
 \hpotential := a + L \nu
$$
provides the desired tensor field.
\hfill $\Box$ \bigskip

Next we consider $t_{ij} \in S_0^2$ with $\mathrm{rot}_1 \,\mathrm{div}_2 \,t = 0$. We will show there exist $v \in C^\infty$ and $\chi_{ij} \in S_0^2$ so that $t = \Koperator  \chi + L d \,v$:

\begin{Theorem}
\label{T13XII19.2}
Let $(M,g)$ be a three-dimensional Einstein space.
On any open subset $\Omega$  of $M$ such that $H_1(\Omega)=\{0\}=H_2(\Omega)$,
%\blue{the right-end of the momentum complex \eqref{mom}
% is
% exact. In other words,}
  a trace-free and symmetric tensor field $t_{ij}$ satisfies
\begin{equation}\label{30XII19.3}
  \mathrm{rot}_1 \,\mathrm{div}_2 \,t = 0
%  \,.
\end{equation}
if and only if
there exists a symmetric trace-free two-covariant tensor field $\tpotential $ and a function $V$ on $\Omega$  such that
%\ptcr{no control about boundary behaviour, right?} \bb{right}
%
\begin{equation}\label{30XII19.2}
\Koperator   \tpotential + L dv  = t
  \,.
\end{equation}
\end{Theorem}

{\noindent\sc Proof.}
The necessity has been established in Section~\ref{ss10I20.1}. For sufficiency, note first
that there exists a function $\tau$ so that $D_i \tau = D^j t_{ij}$. Thus $t_{ij} - \tau g_{ij}$ is divergence-free, whence is $G_i = (t_{ij} - \tau g_{ij}) \xi^j$ for every Killing vector $\xi$.
An argument similar to the one leading to \eqref{10I20.11} shows that
there exists $G_{ij} = G_{[ij]}$ such that $G_i = D^j G_{ij}$ and
%and this $G_{ij}$ has to be  of the form
%
\begin{equation}\label{insert}
G_{ij} = U_{ijk} \xi^k + V_{ijk\ell} D^k \xi^\ell
\end{equation}
with smooth fields $U_{ijk} = U_{[ij]k}$ and $V_{ijk\ell} = V_{[ij][k\ell]}$ and, of course, $D_{(i} \xi_{j)} = 0$. As is well known, the last equation implies that
\begin{equation}\label{kill}
D_i D_j \xi_k = - \frac{R}{3} g_{i[j} \xi_{k]}
 \,.
\end{equation}
Inserting (\ref{insert}) into $G_i = D^j G_{ij}$ and using (\ref{kill}) we find the two equations
\begin{equation}\label{one}
U_{i[jk]} + D^m V_{imjk} = 0
 \,,
\end{equation}
and
\begin{equation}\label{two}
D^
\ell U_{i\ell k} + \frac{R}{3} V_{\ell i}{}^\ell{}_k = t_{ik} - \tau g_{ik}
 \,.
\end{equation}
Using the identity
\begin{equation}
U_{ijk} = U_{i[jk]} + U_{j[ki]} - U_{k[ij]}
\end{equation}
and inserting (\ref{one}) into (\ref{two}) there follows
\begin{equation}
t_{ij} = \tau g_{ij} + \frac{R}{3} V_{\ell i}{}^\ell{}_j - D^\ell D^m (V_{im \ell j} + V_{\ell mji} - V_{jmi\ell})
 \,.
  \label{30XII19.5}
\end{equation}
Using the Ricci identity in the next-to-last term we find after some calculation that
\begin{equation}\label{con}
t_{ij} = \tau g_{ij} + \frac{R}{3} V_{ij} - 2 D^\ell D^m V_{(i|m\ell|j)}\,,
\end{equation}
where $V_{ij} = V_{m(i}{}^m{}_{j)}$.

Now, we have the trivial decomposition
\begin{equation}\label{besse1}
V_{im\ell j} = \frac{1}{2} (V_{im\ell j} - V_{\ell jim}) +  \frac{1}{2} (V_{im\ell j} + V_{\ell jim})\,,
\end{equation}
where the first part is antisymmetric under pair-interchange. For the second,  symmetric part we have (compare~\cite[Definition~1.107]{Besse})
\begin{equation}\label{besse2}
 \frac{1}{2} (V_{im\ell j} + V_{\ell jim}) =   V_{[im\ell j]} + r_{im\ell j}\,.
\end{equation}
where $r_{imlj}$ is an algebraic Riemann tensor.
It follows
\begin{equation}\label{besse}
V_{im\ell j} = \frac{1}{2} (V_{im\ell j} - V_{\ell jim}) + V_{[im\ell j]} + r_{im\ell j}\,.
\end{equation}
Now, consider the first two terms (within the round brackets) on the right-hand side of (\ref{besse}). They can   be rewritten as
\begin{equation}
- 2 D^{[\ell} D^{m]} V_{(i|[m\ell]|j)}
 \,,
\end{equation}
and give no contribution to (\ref{con})  for no obvious reason. This can be seen as follows:
\begin{eqnarray}
 \nonumber
2 D^{[\ell} D^{m]} V_{i[\ell m]j} & = &
 R^{\ell m}{}_i{}^k V_{k\ell mj} + R^{\ell m}{}_j{}^k V_{i\ell mk} = \frac{R}{3}(g_{i[\ell} g_{m]k}  V^{k\ell m}{}_j + g_{j[\ell} g_{m]k} V_i{}^{\ell mk})
 \\
 & = &
   \frac{R}{6}(V_{mi}{}^m{}_j - V_i{}^m{}_{jm}) = 0
    \,.
\end{eqnarray}
Next, in dimension three, the before-last term in (\ref{besse}) vanishes. The last term in (\ref{besse}), again in three space dimensions, is of the form
\begin{equation}\label{riem}
r_{i\ell mj} = 2 g_{m[i} V_{\ell]j} - 2 g_{j[i} V_{\ell]m} - V_n{}^n g_{m[i} g_{\ell]j} = r_{mji\ell}
 \,.
\end{equation}
In order to determine its contribution to (\ref{con}),
we calculate as follows
 \ptcheck{3II120}
\begin{multline}\label{notrace}
- 2 D^m D^\ell (2 g_{m[i} V_{\ell]j} - 2 g_{j[i} V_{\ell]m} - V_\ell{}^\ell g_{m[i} g_{\ell]j}) = 2 \Delta V_{ij} - 4 D_{(i} D^\ell V_{j)\ell} + D_i D_j V_\ell{}^\ell \\
 -   R  V_{ij} + 2 D^\ell D^m V_{\ell m} g_{ij} - \Delta V_l{}^\ell g_{ij} + \frac{R}{3} V_\ell{}^\ell g_{ij}
 \,.
\end{multline}
Note that this is symmetric in $i$ and $j$, which takes care of the symmetrisation occurring in (\ref{con}). Inserting (\ref{notrace}) into (\ref{con}) and taking a trace yields
\begin{equation}\label{trace}
\tau = - \frac{2}{3} D^\ell D^m V_{\ell m} - \frac{R}{9} V_\ell{}^\ell
\,,
\end{equation}
so that
 \ptcheck{3II120}
\begin{equation}\label{finale}
t_{ij} = 2 \Delta V_{ij} - 4 D_{(i} D^\ell V_{j)\ell} + D_i D_j V_\ell{}^\ell - \frac{2 R}{3}(V_{ij} - \frac{1}{3} g_{ij} V_\ell{}^\ell) + \frac{4}{3} D^\ell D^m V_{\ell m} g_{ij} - \Delta V_\ell{}^\ell g_{ij}
 \,.
\end{equation}
Finally, setting $- \frac{1}{2} V_{ij} = \tpotential_{ij} - \frac{1}{6} g_{ij} V_n{}^n$ and $- \frac{1}{3} V_n{}^n = 2 v$,
%and taking a trace of (\ref{finale})
we get (see (\ref{added}))
 \ptcheck{3II120}
\begin{equation}\label{KL}
t_{ij} = (\Koperator  \tpotential)_{ij} + (L d \,v)_{ij}
 \,.
\end{equation}
as promised.
\hfill $\Box$
\ptcheck{24XII19}

\medskip

We have the following variation of Theorem~\ref{T13XII19.2}:

\begin{corollary}
\label{C13XII19.3}
Under the conditions of Theorem~\ref{T13XII19.2},
a symmetric tensor field $K_{ij}$ on $\Omega$ satisfies (\ref{18I20.1}) if and only if
there exists a symmetric two-covariant tensor field $\psi_{ij}$ on $\Omega$  such that \request{
$K$ is given by \eqref{18I20.2}.}
%\begin{equation}
%\label{18I20.2}
%  K = \check  \Koperator \kpotential + \frac{1}{3 }\tr_g K g
%  \,,
%\end{equation}
%
%where $\check \Koperator$ is defined by (\ref{added2}) and $\tr_g K$ by (\ref{24I20.1})
\end{corollary}

\proof
In the notation of the proof of Theorem~\ref{T13XII19.2},
set $\psi_{ij} := - \frac 12 V_{ij}$, $\tr_g K := \frac{3 \tau}{2}$ (see \eqref{16XII19.31}) and use
(\ref{trace})-(\ref{finale}).
\hfill $\Box$
%

%\ptc{some general notes momentarily commmented out}
%\input{LinearisedEvolutionWithLambda}
%\input{LinearisedEvolutionWithLambdaPart1Or2NotSureWhich}

%\ptc{obsolete cut and paste from the previous paper commented out; Weyl section adapted to the current case}
%\input{Shielding}
%\input{WeylFormulation}

\appendix

\appendix
\section{Ellipticity}
\label{A9I20.1}

In this appendix we verify the ellipticity, or lack thereof, of the complexes discussed in Section~\ref{ss10I20.1}. For this we need to calculate the symbols of the operators involved, and determine the relevant images and kernels.

Given an operator $A$ we denote by $\sigma_k(A)$ its symbol, where $k=(k_i)$ is the covector argument.
The tensor field $h_{ij}$ is assumed to be symmetric;   we will \emph{not} assume that $h_{ij}$ is traceless here, but tracelessness will  be assumed in our applications.
We have:
\begin{eqnarray}
  \red{\sigma_k}(d)(f)  &=& (f k_i)
  \,,
\\
  \red{\sigma_k}(L)(Y) &=& (k_i Y_j + k_j Y_i - \frac{2}{3 }k_\ell Y^\ell g_{ij})
  \,,
\\
  \red{\sigma_k}(\mathrm{div}_1)(Y) &=& (k_iY^i)
  \,,
\\
  \red{\sigma_k}(\mathrm{rot}_2)(h) &=& (\epsilon_{i\ell m } k^\ell  h^{ m} {}_j +
  \epsilon_{j\ell m }  k^\ell  h^{ m} {}_i )
  \,,
\\
  \red{\sigma_k}(\mathrm{div}_2)(h) &=& (k_i h^{ij} )
  \,.
\end{eqnarray}
This implies
\begin{eqnarray}
\lefteqn{
  \red{\sigma_k}(\mathrm{div}_1 \mathrm{div}_2)(h)=
  (k_i k_j h^{ij} )
   \,,
   }
   &&
\\
\lefteqn{
  \red{\sigma_k}(\mathrm{rot}_1 \mathrm{div}_2)(h)=
   \big(
    \epsilon_{ij\ell  } k^j k_m h^{\ell m}
     \big)
   \,,
   }
   &&
\\
\lefteqn{
  \red{\sigma_k}( L d)(f)=
   \big(
    2 \big(k_i k_ {j}   - \frac{1}{3 }|k|^2 g_{ij}\big)f
     \big)
   \,,
   }
   &&
\\
\lefteqn{
  \red{\sigma_k}( L \mathrm{rot}_1 )(Y)=
   \big(
    k_i  \epsilon_{j\ell m } k^\ell Y^m +
     k_j  \epsilon_{i\ell m } k^\ell Y^m
     \big)
   \,,
   }
   &&
\\
\lefteqn{
  \red{\sigma_k}( L \mathrm{div}_2 )(h)=
   \big(
    k_i k_\ell h^{\ell}{}_{j} + k_j k_\ell h^{\ell}{}_{i} - \frac{2}{3 }k^\ell k^m h_{\ell m} g_{ij}
     \big)
   \,,
   }
   &&
\\
\lefteqn{
  \red{\sigma_k}((\mathrm{rot}_2)^2)(h)=
   }
   &&
   \nonumber
\\
 &&
   \big(
    \epsilon_{i\ell m } k^\ell
   (\epsilon^{ mnp} k_n h_{jp} +\epsilon_j{}^{ np} k_n h^{m}{}_{p} ) +
  \epsilon_{j\ell m }  k^\ell
   (\epsilon^{ mnp} k_n h_{ip} +\epsilon_i{}^{  np} k_n h^{m{}}{}_{p} )
   \big)
   \nonumber
\\ &=&
   \big(k_n  k^\ell(
    \epsilon_{i\ell m }
    \epsilon^{ mnp}h_{jp} +
  \epsilon_{j\ell m }
    \epsilon^{ mnp} h_{ip}    )
%\\
% &&
+
   k_n  k^\ell(\epsilon_{i\ell m }\epsilon_j{}^{ np}+
    \epsilon_{j\ell m }  \epsilon_i{}^{  np})   h^{m{}}{}_{p}
      \big)
       \nonumber
\\
 &=&
 \big( k^\ell ( k_i h_{j\ell} +
  k_j h_{i\ell}    )   - 2 |k|^2 h_{ij}
%   \nonumber
%\\
% &&
+
   (\epsilon_{i\ell m }\epsilon_j{}^{ np}+
    \epsilon_{j\ell m }  \epsilon_i{}^{  np}) k^\ell k_n h^{m{}}{}_{p} \big)
       \nonumber
\\
 &=&
 \Big( 3 k^\ell ( k_i h_{j\ell} +
  k_j h_{i\ell}    )
   - 4 |k|^2 h_{ij}
   +2 \big( (|k|^2 g_{ij} - k_i k_j) h^\ell{}_\ell
   -
   k^\ell k^m h_{\ell m} g_{ij} \big)
   \Big)
  \,,
       \nonumber
\\
 &&
\\
\lefteqn{
 \red{\sigma_k}(K)(h)= \red{\sigma_k}((\mathrm{rot}_2)^2)(h)
   +  \red{\sigma_k}( L \mathrm{div}_2 )(h)
   }
   &&
       \nonumber
\\
 &=&
 2\Big( 2 k^\ell ( k_i h_{j\ell} +
  k_j h_{i\ell}    ) - \frac{4}{3}
   k^\ell k^m h_{\ell m} g_{ij}
   - 2 |k|^2 h_{ij}
       \nonumber
\\
 &&
   +    (|k|^2 g_{ij} - k_i k_j) h^\ell{}_\ell
   \Big)
  \,,
\\
\lefteqn{
  \red{\sigma_k}(H)(h)= \red{\sigma_k}((\mathrm{rot}_2 K )(h)
   }
   &&
   \nonumber
\\
 &=& 4\Big(
  k^n k^\ell ( k_i \epsilon_{j\ell m}+ k_j \epsilon_{i\ell m} ) h^m{}_n
   - k^\ell |k|^2 ( \epsilon_{i\ell m} h^{m}{}_{j} + \epsilon_{j\ell m} h^m{}_i)
   \Big)
  \,.
\end{eqnarray}
\ptcheck{24XII19}
To check ellipticity of the complexes under consideration we need to calculate the following images $\Ima \red{\sigma_k}$ and  kernels $\Ker \red{\sigma_k}$, for $k\ne 0$. By homogeneity and rotation invariance  it suffices to analyse the case where the coordinate differentials $dx^i$ form an orthonormal basis with $k_idx^i = dx^1$. Letting the indices $A,B$ run over $\{2,3\}$ we then have
\ptcheck{24XII19, some corrections}
\begin{eqnarray}
\lefteqn{
  \Ima \red{\sigma_k}( d) %(f)
   =
\Span{dx^1 }
   =
\Span{k}
   \,,
   }
   &&
\\
\lefteqn{
  \Ima \red{\sigma_k}(\mathrm{div}_1 )
   =
\R
%  (k_i k_j h^{ij} )
   \,,
   }
   &&
\\
\lefteqn{
  \Ima \red{\sigma_k}(\mathrm{div}_1 \mathrm{div}_2) %(h)
   = \R
%\Span{(dx^1)^2}
%  (k_i k_j h^{ij} )
   \,,
   }
   &&
\\
\lefteqn{
  \Ima \red{\sigma_k}(\mathrm{rot}_1)
   =
\Span{dx^A}
   =
{k^\perp}
   \,,
   }
   &&
\\
\lefteqn{
  \Ima \red{\sigma_k}(\mathrm{rot}_1 \mathrm{div}_2) %(h)
   =
\Span{ dx^A}
   =
{k^\perp}
%   \big(
%    \epsilon_{ij\ell  } k^j k_m h^{\ell m}
%     \big)
   \,,
   }
   &&
\\
 \label{30XII19.31}
\lefteqn{
  \Ima \red{\sigma_k}(\mathrm{rot}_2)
   =
\Span{dx^1 dx^B,dx^2 dx^3, (dx^2)^2-(dx^3)^2}
%  (k_i k_j h^{ij} )
   \,,
   }
   &&
\\
\lefteqn{
  \Ima \red{\sigma_k}( L d)%(f)
   =
\Span{2 (dx^1)^2 - (dx^2)^2 - (dx^3)^2}
%    \big(
%    2 \big(k_i k_ {j}   - \frac{1}{3 }|k|^2 g_{ij}\big)f
%     \big)
   \,,
   }
   &&
\\
\lefteqn{
  \Ima \red{\sigma_k}( L \mathrm{rot}_1 )%(Y)
   =
\Span{dx^1 dx^A}
%    \big(
%    k_i  \epsilon_{j\ell m } k^\ell Y^m +
%     k_j  \epsilon_{i\ell m } k^\ell Y^m
%     \big)
   \,,
   }
   &&
\\
\lefteqn{
  \Ima \red{\sigma_k}(K)%(h)
   =
       \Span{2(dx^1)^2 - (dx^2)^2 - (dx^3)^2\,,\, dx^2 dx^3\,,\, (dx^2)^2 -(dx^3)^2
   }
   }
   &&
   \nonumber
\\
 &&
    \oplus\,
     \Span{(dx^2)^2  +(dx^3)^2}
   \,,
    \phantom{xxxxxxxxxxxxxxxxxxxxxxxxxxxxxxxx}
       \label{20XII19.1}
\\
\lefteqn{
  \Ima \red{\sigma_k}(H)(h)=
   \Span{dx^2 dx^3\,,\, (dx^2)^2 - (dx^3)^2}
    \,,
   }
%   &&
%   \nonumber
%\\
% &=& 4\Big(
%  k^n k^\ell ( k_i \epsilon_{j\ell m}+ k_j \epsilon_{i\ell m} ) h^m{}_n
%   - k^\ell |k|^2 ( \epsilon_{i\ell m} h^{m}{}_{j} + \epsilon_{j\ell m} h^m{}_i)
%   \Big)
%\\
% &=&
% 2\Big( 2 k^\ell ( k_i h_{j\ell} +
%  k_j h_{i\ell}    ) - \frac{4}{3}
%   k^\ell k^m h_{\ell m} g_{ij}
%   - 2 |k|^2 h_{ij}
%       \nonumber
%\\
% &&
%   +    (|k|^2 g_{ij} - k_i k_j) h^\ell{}_\ell
%   \Big)
%  \,,
%\\
%\lefteqn{
%  \Ima \red{\sigma_k}(H)(h)=
%   }
%   &&
%   \nonumber
%\\
% &=& 4\Big(
%  k^n k^\ell ( k_i \epsilon_{j\ell m}+ k_j \epsilon_{i\ell m} ) h^m{}_n
%   - k^\ell |k|^2 ( \epsilon_{i\ell m} h^{m}{}_{j} + \epsilon_{j\ell m} h^m{}_i)
%   \Big)
%  \,.
\end{eqnarray}
with the last factor in \eqref{20XII19.1} absent if $h$ is traceless. Furthermore
%\ptcr{24XII19 , but: \eqref{30XII19.12} checked;  \eqref{30XII19.14} corrected;  \eqref{30XII19.31} checked, and also I  confirm  \eqref{30XII19.32}}
%
\begin{eqnarray}
 \label{30XII19.11}
\lefteqn{
  \Ker \red{\sigma_k}(\mathrm{div}_1 )
  =
 \Span{dx^A}
   =
 {k^\perp} =  \Ima \red{\sigma_k}(\mathrm{rot}_1)
 = \Ima \red{\sigma_k}(\mathrm{rot}_1 \mathrm{div}_2)
%  (k_i k_j h^{ij} )
   \,,
   }
   &&
\\
 \label{30XII19.12}
\lefteqn{
  \Ker \red{\sigma_k}(\mathrm{div}_2 )
  =
 \Span{dx^A dx^B}
  =|_{\tr_gh=0} \,
  \Ima \red{\sigma_k}(H  )
   \,,
   }
   &&
\\
 \label{30XII19.12}
\lefteqn{
  \Ker \red{\sigma_k}(\mathrm{div}_1 \mathrm{div}_2) %(h)
   =
\Span{dx^1 dx^A, dx^A dx^B}
%  (k_i k_j h^{ij} )
   \,,
   }
   &&
\\
 \label{30XII19.13}
\lefteqn{
  \Ker \red{\sigma_k}(\mathrm{rot}_1)
   =
\Span{dx^1} =
\Span{k}
 =   \Ima \red{\sigma_k}( d)
   \,,
   }
   &&
\\
 \label{30XII19.14}
\lefteqn{
  \Ker \red{\sigma_k}(\mathrm{rot}_1 \mathrm{div}_2) %(h)
   =
\Span{ (dx^1)^2\,,\, dx^Adx^B}
% =|_{\tr_gh=0}  \Ima \red{\sigma_k}( \mathrm{rot}_2 )
%   \big(
%    \epsilon_{ij\ell  } k^j k_m h^{\ell m}
%     \big)
   \,,
   }
   &&
\\
 \label{30XII19.15}
\lefteqn{
  \Ker \red{\sigma_k}(\mathrm{rot}_2)
   =
 \Span{(dx^1)^2,  (dx^2)^2+(dx^3)^2}
   }
   &&
% \nonumber
\\
 \label{30XII19.16}
 &&
 =|_{\tr_gh=0} \,
\Span{2(dx^1)^2 - (dx^2)^2-(dx^3)^2}
 \nonumber
\\
 &&
 =|_{\tr_gh=0} \,
  \Ima \red{\sigma_k}( L d)
   \,,
    \phantom{xxxxxxxxxxxxxxxxxxxxxxxxxxxxxxxx}
 \label{30XII19.32}
\\
 \label{30XII19.17}
\lefteqn{
  \Ker \red{\sigma_k}( L d)%(f)
   =
    \{0\}
%    \big(
%    2 \big(k_i k_ {j}   - \frac{1}{3 }|k|^2 g_{ij}\big)f
%     \big)
   \,,
   }
   &&
\\
 \label{30XII19.18}
\lefteqn{
  \Ker \red{\sigma_k}( L \mathrm{rot}_1 )%(Y)
   =
\Span{ dx^1} =\Span{k}
=  \Ima \red{\sigma_k}(d)
%    \big(
%    k_i  \epsilon_{j\ell m } k^\ell Y^m +
%     k_j  \epsilon_{i\ell m } k^\ell Y^m
%     \big)
   \,,
   }
   &&
\\
\lefteqn{
  \Ker \red{\sigma_k}(K)%(h)
   =
       \Span{  dx^1 dx^A
   }
       =
  \Ima \red{\sigma_k}( L \mathrm{rot}_1 )
   \,,
   }
   &&
       \label{20XII19.3}
\\
\lefteqn{
  \Ker \red{\sigma_k}(H)
   =
   \Span{dx^1 dx^A\,,\, (dx^1)^2\,,\,(dx^2)^2+ (dx^3)^2}
    \,,
   }
    \label{21XII19.1}
\end{eqnarray}
where we write $
 =|_{\tr_gh=0}$ for the equality when ${\tr_gh=0}$.
 \ptcheck{24XII19}

\bigskip

\noindent{\sc Acknowledgements:}
We are very grateful to Olaf Kr\"uger for typesetting Figure~\ref{F19XII19.1}. The research of PTC  was
supported
by the Austrian Research Fund (FWF), Project  P29517-N27, and
by the Polish National Center of Science (NCN) under grant 2016/21/B/ST1/00940.

\providecommand{\bysame}{\leavevmode\hbox to3em{\hrulefill}\thinspace}
\providecommand{\MR}{\relax\ifhmode\unskip\space\fi MR }
\providecommand{\MRhref}[2]{%
  \href{http://www.ams.org/mathscinet-getitem?mr=#1}{#2}
}
\providecommand{\href}[2]{#2}

\end{document}